\documentclass[preprint,12pt]{elsarticle}

\usepackage[numbers]{natbib}
\usepackage{amsmath,amssymb,amsfonts,dsfont}
\usepackage{algorithmic}
\usepackage{graphicx}
\usepackage{textcomp}
\usepackage{caption}
\usepackage{color,soul}
\usepackage{url}
\usepackage{subfigure}
\usepackage{epstopdf}
\usepackage{multirow}
\usepackage{bm}
\usepackage{ulem}
\usepackage{cancel}
\begin{document}

\begin{frontmatter}



\title{Interlayer link prediction in multiplex social networks: an iterative degree penalty algorithm}


\author{Rui Tang$^{1}$, Shuyu Jiang$^{1}$, Xingshu Chen$^{1,2*}$, Haizhou Wang$^{1,2}$, Wenxian Wang$^{2}$, and Wei Wang$^{2*}$}
\cortext[cor1]{Corresponding Author: chenxsh@scu.edu.cn, wwzqbx@hotmail.com}

\address{1. College of Cybersecurity, Sichuan University,
Chengdu 610065, China}
\address{2. Cybersecurity Research Institute, Sichuan University,
Chengdu 610065, China}

\begin{abstract}
Online social network (OSN) applications provide different experiences; for example, posting a short text on Twitter and sharing photographs on Instagram. Multiple OSNs constitute a multiplex network. For privacy protection and usage purposes, accounts belonging to the same user in different OSNs may have different usernames, photographs, and introductions. Interlayer link prediction in multiplex network aims at identifying whether the accounts in different OSNs belong to the same person, which can aid in tasks including cybercriminal behavior modeling and customer interest analysis. Many real-world OSNs exhibit a scale-free degree distribution; thus, neighbors with different degrees may exert different influences on the node matching degrees across different OSNs. We developed an iterative degree penalty (IDP) algorithm for interlayer link prediction in the multiplex network. First, we proposed a degree penalty principle that assigns a greater weight to a common matched neighbor with fewer connections. Second, we applied node adjacency matrix multiplication for efficiently obtaining the matching degree of all unmatched node pairs. Thereafter, we used the approved maximum value method to obtain the interlayer link prediction results from the matching degree matrix. Finally, the prediction results were inserted into the priori interlayer node pair set and the above processes were performed iteratively until all unmatched nodes in one layer were matched or all matching degrees of the unmatched node pairs were equal to 0. Experiments demonstrated that our advanced IDP algorithm significantly outperforms current network structure-based methods when the multiplex network average degree and node overlapping rate are low.

\end{abstract}


\end{frontmatter}

% \linenumbers

%\tableofcontents

\section{Introduction}
With rapid developments in Internet technology, Internet surfing has become increasingly convenient and efficient in recent years. Online social network (OSN) applications, such as Twitter, Facebook, Instagram, and LinkedIn, have rapidly been integrated into people's everyday lives and have become major social communication tools. These multiple OSNs constitute a multiplex network, where each OSN can be represented as a layer within the multiplex network. In this multiplex network, nodes represent user accounts and intralayer links capture friendships or follower-followee relations. If two accounts in different OSNs belong to the same user, an interlayer link will exist between the nodes. The various OSN applications provide users with different functional experiences. For example, people generally use LinkedIn to follow work-related contents, post short text descriptions on Twitter to express their experiences at the time, and share their photographs on Instagram. Analyzing these OSNs has a significant impact on society, politics, economy, etc~\cite{ShiChuan2017,ShuKai2017,fujita2019hypotheses,capuano2017fuzzy}.

In general, the OSN can be characterized by a complex network, in which nodes represent individuals and links capture the relationships between them~\cite{Newman2010}. Several researchers have conducted studies in this field relating to clustering~\cite{HanXiaohui2017}, link prediction~\cite{chi2019link}, information diffusion~\cite{ShiChuan2017}, and community detection~\cite{yin2017approach, zhao2019incremental}, among others. However, it is difficult to apply such a representation to describe multilayer structures such as multiple OSNs, multiple transportation networks~\cite{Boccaletti2014-multinet} as well as the dynamic networks~\cite{chi2019link, zhao2019incremental}. The multilayer structures have a significant influence on the aspects of cascade~\cite{Gao2011, Gao2011a}, propagation~\cite{GranellClara2013, WangWei2014, WangWei2016, Wangwei2019}, synchronization~\cite{ZhangXiyun2015}, and game~\cite{WangZhen2013, WangZhen2013a, WangZhen2014}. In recent years, the multiplex network~\cite{Gao2011, Kivela2014, Emanuele2018-MultiNetwork} has emerged to characterize these multilayer structures. It explicitly incorporates multiple connectivity channels into a system and provides a natural description for systems in which entities have a different set of neighbors in each layer~\cite{Domenico2013}.

 A user may always be active in different OSNs using different accounts; in most cases, it is not known  whether accounts across different OSNs belong to the same person. This means that most of the interlayer links are not given in advance. To identify whether accounts in different OSNs belong to the same person, the structure or feature information should be leveraged to predict the interlayer links, which is a challenging problem in network analysis in recent years~\cite{ShuKai2017}. It has been established that no OSN application can completely replace all other similar software at present. Therefore, online users generally register multiple accounts for using these OSNs. When one person registers his or her accounts in different OSNs, he or she may use different usernames, photographs, and introductions for the purpose of privacy protection or anonymity. In this manner, users can confidently use the Internet to chat, make friends, and share information on different social media, according to their personal preferences. However, anonymity also poses harms to society to a certain extent.
 For example, criminals may register large quantities of different accounts on OSN applications. They subsequently engage in illegal activities, such as spreading rumors, spreading viral links, and inducing financial fraud on these applications~\cite{HanXiaohui2017}. Given effective methods to determine the correspondence between accounts of the same user on different OSNs, we can establish the patterns of criminal violations of laws, model their online behavior, lock their geographical locations, and even determine their real identities, thereby effectively striking against them.

Moreover, numerous other benefits are offered when predicting the interlayer links in the multiplex network consisting of multiple OSNs. For example, business site owners can be aided in studying user behaviors, analyzing the interests of customers, and analyzing the factors that affect their decisions~\cite{LuChunTa2014}. Furthermore, OSN users can be kept up to date with their virtual contacts from different OSNs in an integrated environment~\cite{Vosecky2009}.

Three main approaches are available for interlayer link prediction in the multiplex social network: (i) feature-based interlayer link prediction, (ii) network-based interlayer link prediction, and (iii) a combination of multiple approaches. Among these, network-based methods have been applied extensively in recent years~\cite{WangShaokai2019, ZhangSi2019-www, ChuXiaokai2019-www, WangYongqing2019-www, ZhanQianyi2018, ZhouXiaoping2018, ZhouFan2018, ZhouXiaoping2018-IEEE, ZhangJing2018, ZhangSi2017-ICDM, ZhouXiaoping2016, LiuLi2016, ManTong2016-IJCAI, ZhangSi2016-KDD, ZhangJiawei2015-DM, TanShulong2014-AAAI}. Its strategies for predicting interlayer links rely on examining the structures of social graphs on different OSN platforms~\cite{ZhouFan2018}.

In the process of using network structures for interlayer link prediction, in general, only the attributes of the nodes to be predicted are considered, along with the importance of the attributes of their neighbors, such as ignoring the neighbors' degrees~\cite{ZhouXiaoping2016}. Many real-world OSNs exhibit a scale-free property. Their degree distribution follows a power law distribution, which is common knowledge in numerous real-world networks~\cite{FengRui2018, ZhouTao2011}. The number of common matched neighbors and their degree attributes may have different effects on the matching degree of two nodes in different layers for the prediction of their interlayer link.

In this study, we developed an iterative degree penalty (IDP) algorithm for interlayer link prediction in the multiplex network. The major contributions of this paper can be summarized as follows:
\begin{itemize}
\item We propose a degree penalty principle to calculate the matching degree between the unmatched nodes across different layers in the multiplex network. In the real world, people's friend relationship cycles are highly individual \cite{ZhouXiaoping2016}. Therefore, with more common matched neighbors, a higher probability exists that two unmatched nodes across different layers have an interlayer link. However, common matched neighbors with fewer connections are assigned higher weights for the matching degree between two unmatched nodes. Our degree penalty principle uses the inverse log frequency of the common matched neighbors of two unmatched nodes to calculate the matching degree, thereby effectively addressing the above two situations.

\item We develop an iterative algorithm to determine additional hidden interlayer links. We use node adjacency matrix multiplication for efficiently obtaining the matching degree of all unmatched node pairs. Thereafter, we use the approved maximum value method to obtain the interlayer link prediction results from the matching degree matrix. Finally, the prediction results are inserted into the priori interlayer node pair set, and the above processes are performed iteratively until all of the unmatched nodes in one layer are matched or all the matching degrees of the unmatched node pairs are equal to 0.

\item We verify the effectiveness of our advanced IDP algorithm on both artificial scale-free and real-world networks. The results demonstrate that the IDP algorithm significantly outperforms the existing network structure-based method FRUI: the recall rate increases by a maximum of 36.6\%  and an average of 7.0\%  when the priori interlayer link rates are less than 10\% on the real-world networks.
\end{itemize}

The remainder of this paper is organized as follows. Section 2 reviews related works. Section 3 describes the preliminaries and presents the problem definitions. Section 4 explains the proposed IDP algorithm. Section 5 presents the experimental results. Section 6 concludes the paper and provides future research directions.

\section{Related Works}
The problem of interlayer link prediction in a multiplex network constituted by multiple OSNs has attracted substantial research attention in the past decade~\cite{ShuKai2017}. Existing research efforts can be divided into three categories: feature-based prediction, network-based prediction, and a combination of multiple approaches.

\subsection{Feature-based prediction}
In several works, features have been extracted from profiles and contents. These studies have adopted machine learning techniques to predict the interlayer links across multiple OSNs~\cite{ShuKai2017}. Profile attributes include the username, gender, birthday, address, experience, and image, among others~\cite{ZhaoDongsheng2018}. The username has been found to be the most important attribute in the profile and has been explored extensively. Zafarani and Liu~\cite{Zafarani2009} formalized the problem of mapping among identities across multiple websites, and conducted similar empirical tests on thousands of usernames across 12 social networks, thereby empirically validating several hypotheses. They further developed a supervised methodology for the same problem in Ref. \cite{Zafarani2013}, which extracted behavioral features of usernames based on the priori knowledge of linguistics and human behavior according to three aspects: human limitations, exogenous factors, and endogenous factors. Perito et al. \cite{Perito2011} determined that a significant portion of the users can be linked by means of their usernames, and identified users with binary classifiers. Liu et al. \cite{LiuJing2013-WSDM} differentiated among users with the same usernames and proposed an unsupervised approach to match users.
Among usernames, the recognition of photographs may provide a means of mapping users. Although usernames can be used to predict the interlayer links of the multiplex social network, it is difficult to use them in large-scale situations because some users may have the same username. Acquisti et al. \cite{Acquisti2014} used publicly available photos for large-scale individual re-identification. However, users tend to publish different pieces of information in different OSNs ~\cite{labitzke2011your}. Only a handful of users may put their personal photos on different OSNs. Therefore, using photos for identification is appropriate for these specific users.

Apart from usernames and images, several researchers have focused on considering various profile attributes to improve the prediction performance~\cite{Carmagnola2009, Iofciu2011, AbelFabian2013, Goga2013, Cortis2013, MuXin-KDD2016}.
A heuristic approach based on using the username, e-mail, and birth city attributes was proposed by Carmagnola et al. \cite{Carmagnola2009}.
Iofciu et al.~\cite{Iofciu2011} divided profile information into two types: usernames and tags. They determined that users can be identified by analyzing their tagging, and the performance can be improved by combining tags and usernames.
Mu et al.~\cite{MuXin-KDD2016} explored the concept of latent user space to describe the user profiles in different social platforms, so that two individuals with greater similarities would have closer profiles in the latent user space.
Similar areas were also studied in Refs.~\cite{AbelFabian2013, Goga2013, Cortis2013}. Leveraging more profile attributes could improve the prediction performance effectively, but profile attributes may contain information which are null or are not sufficiently strong to indicate user identity.

Content attributes for individuals can reveal their activities, such as posting, mentioning, and commenting in OSNs~\cite{ShuKai2017}. These properties can capture unique characteristics for the interlayer link prediction of multiplex OSNs.
Zheng et al.~\cite{ZhengRong2006} extracted four types of writing-style features (syntactic, lexical, structural, and content) to identify authorship.
Goga et al.~\cite{Goga2013a} exploited the geo-location, timestamp, and writing style of user posts to identify accounts on different OSNs.
Zafarani and Liu~\cite{Riederer2016} extracted trajectory-based content features to capture the unique footprints of user activities for linking the same user accounts across platforms. Nevertheless, the methods based on content attributes often face the problem of data sparsity, because the amount of location data or posted content of different users varies significantly. These approaches are usually confined to some specific OSN applications, which makes it harder to leveraged them on a general multiplex social network.
\subsection{Network-based prediction}
Personal data available on OSNs are usually anonymized owing to privacy concerns, with users often removing their profile and attribute information or replacing these with fake information~\cite{ZhangJiawei2015-DM}. Therefore, using network structures to predict the interlayer links has been the focus of substantial research.
Narayanan and Shmatikov~\cite{Narayanan2010} assumed that a natural person usually has a similar social network in the virtual world.
Based on this assumption, Zhou et al.~\cite{ZhouXiaoping2016} developed a network structure-based method known as the Friend Relationship-based User Identification (FRUI) algorithm. The FRUI algorithm identifies the same user across different OSNs by leveraging the shared friend cycle, and requires several priori matched pairs. They subsequently proposed an approach that did not require priori matched pairs in Ref.~\cite{ZhouXiaoping2018-IEEE}.
Interlayer link prediction in a multiplex network consisting of multiple OSNs formed the maximum common subgraph problem in Ref.~\cite{ZhuYuanyuan2012}, which maximized the number of standard links to obtain one-to-one mapping. The maximum common subgraph problem is an NP-hard problem, and is therefore hardly used in real scenarios.
Tan et al.~\cite{TanShulong2014-AAAI} leveraged a hypergraph to model user relationships and correlate accounts across different OSNs by projecting the manifolds of two networks onto a commonly embedded space.
Moreover, neighborhood-based network features were used for account alignment in~\cite{ZhangYutao2015}.
Zhang et al.~\cite{ZhangJiawei2015-IJCAI} proposed a joint link fusion algorithm to predict the social links and anchor links of two OSNs simultaneously. This algorithm transferred information relating to social links from one network to another. Its time complexity is approximately $O(n^3)$. Numerous researchers have used network embedding~\cite{TangJian2015}, also known as representation learning \cite{PerozziBryan2014}, to predict interlayer links in recent years.
Liu et al.~\cite{LiuLi2016} adopted a network embedding approach to align users across multiple directed OSNs. Their method embedded two OSNs into a common space to capture the social links of user accounts. Man et al.~\cite{ManTong2016-IJCAI} proposed a supervised model to learn cross-layer mapping for interlayer link prediction, which leveraged network embedding to capture the underlying structural regularities. Zhou et al.~\cite{ZhouXiaoping2018} proposed an unsupervised algorithm for user identification, using network embedding and scalable nearest neighbors. A deep reinforcement learning-based framework was developed in Ref.~\cite{ZhouFan2018}, which could embed the global network structure, thereby achieving higher correlation accuracy than other network embedding methods. Similar studies can also be found in Refs.~\cite{WangShaokai2019, ZhangSi2019-www, ChuXiaokai2019-www, WangYongqing2019-www, ZhouXiaoping2018, ZhangJing2018, ZhangSi2017-ICDM, ZhangSi2016-KDD}. Most of the network embedding-based methods implicitly assume that the input networks are complete without missing edges, and it is not easy to add user attribution into the representation vectors.

\subsection{Combination of multiple approaches}
Other relevant approaches have combined the feature-based and structure-based methods for interlayer link prediction; for example, searching the same username from the friend lists of seed users across Facebook and Twitter~\cite{JainParidhi2013}.
Nunes et al.~\cite{Nunes2012} introduced a binary classification method to classify account pairs as belonging to the same person or not. The feature vectors of the classifier were constructed by profile information, descriptions of interests, and friend lists.
Kong et al.~\cite{KongXiangnan2013} studied two fully aligned OSN datasets collected from Foursquare and Twitter to determine the correspondence between different user accounts. They extracted multiple features and formulated the correspondence problem as a stable matching problem. The network structure features were one type among multiple features.
Zhang et al.~\cite{ZhangHaochen2014} explored a probabilistic approach for mapping individuals, in which the user friend or connection count was used as a feature.
Lu et al.~\cite{LuChunTa2014} introduced a methodology to identify customers between customer accounts, e-commerce sites, and OSN applications. The network structure was used to extract user interest features.

In the majority of combination approaches, the problem of attempting to obtain user profile features for the sake of privacy protection exists. Our study focuses on structure-based prediction. Numerous real-world OSNs exhibit a scale-free degree distribution~\cite{ZhouTao2011}. Common matched neighbors with different degrees may have different influences on the matching degrees of nodes across two OSNs. In this study, we developed an IDP algorithm to address this issue.

\section{Preliminaries and problem}
In this section, we introduce the conception of the multiplex network for representing multiple OSN applications, and describe the problem of interlayer link prediction on the multiplex network. The symbols and notations frequently used in this paper are displayed in Table~\ref{tab:symbol}. We use bold uppercase letters for matrices, bold lowercase letters for vectors, and lowercase letters for scalars.
\begin{figure*} [t!]
    \centering
    \includegraphics[width=0.8\textwidth]{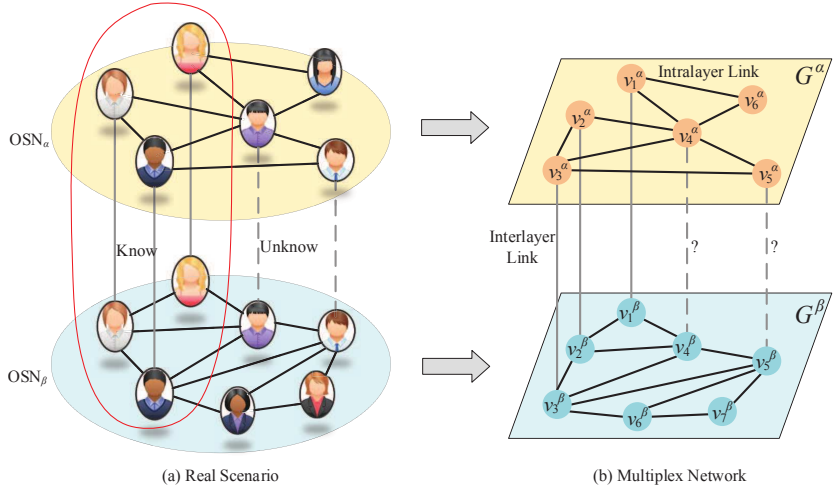}
    \caption{Example of representing relations across multiple OSNs using multiplex network. (a) Real scenario: There are two OSN applications. OSN$_\alpha$ has six user accounts, while OSN$_\beta$ has 7. Each account exhibits certain connections with other accounts on the same OSN. In the red circle, three pairs of accounts belong to three users, and the correspondence among these accounts is revealed in advance. The task is to determine the correspondence of the other accounts.
 (b) Multiplex network: We use a multiplex network $\mathcal{M}$ to represent multiple OSNs. The two OSNs are denoted by layers $\alpha$ and $\beta$. The user accounts are denoted by nodes $v^\alpha_i$ or $v^\beta_j$. The relationships of the accounts connected in an OSN are denoted by interlayer links $e^\alpha_{ij}$ or $e^\beta_{ij}$. The correspondence of two accounts across two OSNs is denoted by the interlayer link $e^{\alpha\beta}_{ij}$. The task of determining the correspondence of accounts across OSNs becomes the prediction of interlayer links in the multiplex network.}
    \label{pic:real_to_multiplx}
\end{figure*}

\subsection{Definitions}
To distinguish the different OSN platforms, we represent the relationship between nodes in one OSN as the graph $G^{\alpha}(V^{\alpha},E^{\alpha})$, where $V^{\alpha}$ and $E^{\alpha}\subseteq V^\alpha \times V^\alpha$ are the sets of nodes and edges of the graph $G^{\alpha}$. The elements of $E^\alpha$ are referred to as intralayer links. The scenario of multiple OSNs is represented by the multiplex network $\mathcal{M}=(g,c)$, where $g={\{ G^\alpha | \alpha \in \{ 1,\cdots,m \} \}} $ is the set of graphs and $ c={\{E^{\alpha \beta}\subseteq V^\alpha \times V^\beta | \alpha,\beta\in\{1,\cdots,m\},\alpha\neq\beta\}} $ is the set of connections between the nodes of graphs $G^\alpha$ and $G^\beta$. If there exists an edge $e^{\alpha\beta}_{ij}$ between node $v^\alpha_i$ in graph $G^\alpha$ and node $v^\beta_j$ in graph $G^\beta$, the accounts represented by the two nodes belong to the same user. The elements of $E^{\alpha \beta}$ are known as interlayer links. Each subnetwork in $g$ is referred to as a layer in multiplex network $\mathcal{M}$. An illustration of multiplex networks is provided in Fig.~\ref{pic:real_to_multiplx}.
\begin{table*}
\caption{Symbols and notations}
\label{tab:symbol}
\centering
\begin{tabular}{p{2cm}p{12cm}}
\hline
%\toprule
\textbf{Symbol }& \textbf{Description}\\
%\midrule
\hline
$\mathcal{M}$ & The multiplex network.\\
$G$ & A social network that is one layer of $\mathcal{M}$. \\
$u,v$  & Nodes in $\mathcal{M}$. \\
$\alpha,\beta$ & Layer indices of $\mathcal{M}$.\\
$e^\alpha,e^\beta$  & Intralayer connections in $G^\alpha$ and $G^\beta$. \\
$\bm{e}$,$\bm{E}$ & Intralayer connection vector and intralayer connection matrix, respectively.\\
$e^{\alpha\beta}$  & Interlayer connection. \\
$i,j,a,b$ & Node indices.\\
$n^\alpha,n^\beta$ & Number of nodes in $G^\alpha$ and $G^\beta$.\\
$n$ & Number of MINPs.\\
 $\Gamma(v_i)$ & Neighbors set of node $v_i$. \\
$r, \bm{R}$ & Matching degree between two interlayer nodes and matching degree matrix for all interlayer nodes, respectively.\\
$k$ & Degree of a node. \\
$\bm{d}$ & Degree vector of all nodes in a layer. \\
$h$& Reciprocal of logarithm for a priori interlayer node degree \\
$\bm{h},\bm{H}$ & Reciprocal vector of logarithm for a priori interlayer node degree and reciprocal matrix of logarithm for all priori interlayer node degrees. \\
$\delta$ & Control parameter for selecting candidate matched interlayer nodes.\\
$\bm{A}$ & Adjacency matrix of a layer.\\
$p$ & The ratio of PINPs to INPs. \\
$s$ & The percentage of remaining nodes when extracting subnetworks to construct the multiplex network. \\
$\Phi$ & Set of PINPs.\\
$\varphi^\alpha,\varphi^\beta$ & Set of priori interlayer nodes in layers $\alpha$ and $\beta$.\\
$\Psi$ & Set of MINPs.\\
$\psi^\alpha,\psi^\beta$ & Set of matched interlayer nodes in layer $\alpha$ and $\beta$.\\
%\bottomrule
\hline
\end{tabular}
\end{table*}
Prior to introducing the details of our proposed method, we present the formal definitions of various important concepts in this paper.

\textbf{Definition 1: Interlayer node pair (INP).} Given a multiplex network $\mathcal{M}$, if an interlayer link exists between a node $v^\alpha_i$ in layer $\alpha$ and a node $v^\beta_j$ in layer $\beta$, we refer to the pair consisting of these two nodes as an INP, which is represented by INP$(v^\alpha_i$ ,$v^\beta_j)$. Moreover, the nodes belonging to the INP are known as interlayer nodes.

\textbf{Definition 2: Matched interlayer node pair (MINP).}
Given a multiplex network $\mathcal{M}$, if a node pair consisting of node $v^\alpha_i$ in layer $\alpha$ and node $v^\beta_j$ in layer $\beta$ is matched as an INP, we refer to this pair as an MINP and denote it by MINP $(v^\alpha_i$ ,$v^\beta_j)$. An MINP is not necessarily an INP, because the results calculated by the algorithm may be incorrect.

\textbf{Definition 3: Priori interlayer node pair (PINP).}
Given a multiplex network $\mathcal{M}$, if several INPs are provided in advance, we refer to these INPs as PINPs. Furthermore, the nodes belonging to the PINP are known as priori interlayer nodes.

\textbf{Definition 4: Unmatched node pair (UNP).}
Given a multiplex network $\mathcal{M}$, if a node $v^\alpha_i$ in layer $\alpha$ and a node $v^\beta_j$ in layer $\beta$ have not been matched, we refer to the pair consisting of these two nodes as a UNP, which is represented by UNP$(v^\alpha_i$ ,$v^\beta_j)$.

\textbf{Definition 5: Common matched neighbor (CMN).}
Given a PINP($v^\alpha_i$,$v^\beta_j$), if node $v^\alpha_a$ in layer $\alpha$ has an intralayer link with node $v^\alpha_i$, and node $v^\beta_b$ in layer $\beta$ has an intralayer link with node $v^\beta_j$, we can state that the PINP($v^\alpha_i$,$v^\beta_j$) is the CMN of nodes $v^\alpha_a$ and $v^\beta_b$.

It is worth noting that the terms graph and layer, link and connection, and interlayer node pair and interlayer link are used interchangeably in this paper.
\subsection{Problem statement}
Supposing that we have a two-layer multiplex network $\mathcal{M}$ with a small set of priori interlayer links, the purpose of the interlayer link prediction problem is to predict which node pairs are most likely to have the interlayer links.

Given a UNP $(u^\alpha_i,u^\beta_j)$ in the multiplex network $\mathcal{M}$, the objective function of the interlayer link prediction can be defined as follows:
\begin{equation}
f(u^\alpha_i,u^\beta_j) = J(r_{ij}) = \left\{
\begin{array}{ll}
1, & \mbox{if $e^{\alpha\beta}_{ij}$ exist,} \\
0, & \mbox{otherwise,}
\end{array}
\right.
\label{eq:objectivefunction}
\end{equation}
where $r_{ij}$ represents the matching degree of the unmatched nodes $v^\alpha_i$ and $v^\beta_j$. The interlayer link prediction problem is converted into the calculation of $r_{ij}$ and the definition of the objective function $J$.

In a real scenario, certain people may possess two or more accounts in the same OSN application. For simplicity, we assume that these multiple accounts belong to different users. This means that the interlayer link prediction problem in this study is a one-to-one matching problem; that is, no two interlayer links share the same node.

\section{IDP algorithm}
In this section, we introduce the iterative interlayer link prediction algorithm, known as the IDP, into the multiplex network to predict the interlayer links based on only a small part of these. Firstly, we propose a degree penalty principle that assigns a greater weight to a CMN with fewer connections. Secondly, we propose node adjacency matrix multiplication for efficiently obtaining the matching degree of all of the UNPs. Thereafter, we use the approved maximum value method to obtain the interlayer link prediction results from the matching degree matrix. Finally, the prediction results are inserted into the PINP set, and the above processes are performed iteratively until all of the unmatched nodes in one layer are matched or all of the match degrees of the UNPs are equal to 0.
\begin{figure*} %[ht!]
%\begin{minipage}[t]{0.45\textwidth}
    \centering
    \includegraphics[width=\textwidth]{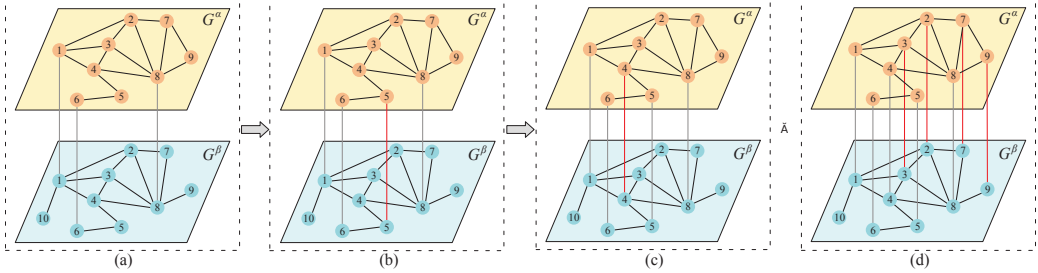}
    \caption{Examples of interlayer link prediction with IDP. (a) A two-layer multiplex network with nine nodes in layer $\alpha$ and ten nodes in layer $\beta$. All of the interlayer links and three of the interlayer links are provided, respectively. Now, we need to use the IDP algorithm to predict the remaining interlayer links. (b) The first iteration to output a prediction result. We use the degree penalty principle to calculate the matching degree of each candidate node pair and set $\delta=1$ to select the maximum pair (5,5) as a result in this iteration. (c) The next iteration for predicting the interlayer links. In this iteration, there are four MINPs. We use the same method to calculate the matching degree, and the result is (4,4). (d) The results in the later iterations, which are (3,3), (2,2), (7,7), and (9,9).}
    \label{pic:IEAA_Procedure}
%\end{minipage}
\end{figure*}

\subsection{Degree penalty principle}
Researchers have established that numerous natural and artificial networks exhibit certain common topological characteristics, such as small world, scale free, and core periphery~\cite{Newman2003}. The scale-free property is valuable in that the degree distribution of a scale-free network follows a power law. Zhou et al. analyzed four leading OSN sites (Delicious, Flickr, Twitter, and YouTube), and found that these sites all exhibit scale-free degree distributions~\cite{ZhouTao2011}.
This means that there exist large nodes with a small degree, as well as several nodes with a large degree. The CMNs with different degrees will have different weights for the UNP matching degrees. For example, if a person has only one friend and he or she follows an account $v^\alpha$ on OSN $G^\alpha$ and an account $v^\beta$ on OSN $G^\beta$, it is highly likely that $v^\alpha$ and $v^\beta$ belong to the same person, namely his or her friend. In contrast, if this person has many friends, it is difficult to determine whether or not the accounts that he or she follows on different OSNs belong to the same person.

Based on the above statements, we propose the principle of the degree penalty of CMNs for calculating the matching degree of the unmatched nodes, which assigns more weights to smaller-degree CMNs. We define
\begin{equation}
r_{ij}=\sum_{\substack{\forall(v^\alpha_a,v^\beta_b)\in \Phi, \\ v^\alpha_a \in\Gamma(u^\alpha_i),\\v^\beta_b\in\Gamma(u^\beta_j)}} \log^{-1}(k_{v^\alpha_a}+1)+\log^{-1}(k_{v^\beta_b}+1)
\label{eq:EAAI}
\end{equation}

as the matching degree of UNP $(u^\alpha_i,u^\beta_j)$. In Eq.~(\ref{eq:EAAI}), $\Phi$ represents the set of PINPs, $\Gamma(u^\alpha_i)$ and $\Gamma(u^\beta_j)$ represent the neighbor sets of nodes $u^\alpha_i$ and $u^\beta_j$, respectively, $k$ represents the node degree, and the constraints in the equation indicate that the PINP $(v^\alpha_a,v^\beta_b)$ is the CMN of UNP $(u^\alpha_i,u^\beta_j)$. It is noteworthy that, if $k_{v^\alpha_a}=1$, $\log(k_{v^\alpha_a})$ will be equal to 0, and $\log^{-1}(k_{v^\alpha_a})$ becomes meaningless. To overcome this problem, we add 1 to each log by means of Laplace smoothing.

The degree penalty principle has the following characteristics:

(i) It can reflect the contribution of the number of CMNs to the matching degree. For any two unmatched nodes $u^\alpha_i$ and $u^\beta_j$, a greater number of CMNs results in their matching degree $r_{ij}$ being larger.

(ii) It can reflect the influence of the degree of CMNs on the matching degree. For any CMNs, a smaller degree results in the weights of these CMNs making a greater contribution to the matching degree of the UNPs connected to them.
\subsection{Calculation of matching degree}
The matching degree of each UNP can be calculated by the degree penalty principle proposed above. Given all of the nodes, interlayer links, and priori interlayer links, the question is how to obtain all matching degrees of the UNPs efficiently. We propose an approach based on matrix operation, inspired by Ref.~\cite{Katz1953}.

Given a multiplex network $\mathcal{M}$, $n^\alpha$ and $n^\beta$ are denoted as the number of nodes in layers $\alpha$ and $\beta$, respectively, while $n$ is the number of PINPs. The number of unmatched nodes in layer $\alpha$ can be represented by $n^\alpha-n$, while the number of unmatched nodes in layer $\beta$ can be represented by $n^\beta-n$.

For a node $v^\alpha_a \in \varphi^\alpha$, where $\varphi^\alpha$ represents the set of priori interlayer nodes in layer $\alpha$, the degree of $v^\alpha_a$ can be expressed as
\begin{equation}
k_{v^\alpha_a}=\sum_{b=1}^{n_\alpha}{e^\alpha_{ab}},
\end{equation}
where $e^\alpha_{ab}$ is equal to 1 if an intralayer link exists between nodes $v^\alpha_a$ and $v^\beta_b$, and 0 otherwise.

By $h^\alpha_a=\log^{-1}(k_{v^\alpha_a}+1)$, we can rewrite Eq.~(\ref{eq:EAAI}) as
\begin{equation}
r_{ij}=\sum_{\substack{\forall(v^\alpha_a,v^\beta_b)\in \Phi, \\ v^\alpha_a \in\Gamma(u^\alpha_i),\\v^\beta_b\in\Gamma(u^\beta_j)}} h^\alpha_a + h^\beta_b.
\label{eq:EAAI_abbr}
\end{equation}

This equation will facilitate the subsequent introduction of matrix operations.

It is clear that, if PINP $(v^\alpha_a,v^\beta_b)$ is the CMN of UNP $(u^\alpha_i,u^\beta_j)$, $e^\alpha_{ia}$ and $e^\beta_{jb}$ will be equal to 1; thus, $e^\alpha_{ia} \cdot h^\alpha_a  \cdot e^\beta_{jb}=h^\alpha_a$ and $e^\alpha_{ia} \cdot h^\beta_b  \cdot e^\beta_{jb} =h^\beta_b$. In contrast, if PINP $(v^\alpha_a,v^\beta_b)$ is not the CMN of UNP $(u^\alpha_i,u^\beta_j)$, $e^\alpha_{ia}$ or $e^\beta_{jb}$ will be equal to 0; thus, $e^\alpha_{ia} \cdot h^\alpha_a \cdot e^\beta_{jb}=0$ and $e^\alpha_{ia} \cdot h^\beta_b \cdot e^\beta_{jb}=0$. Therefore, Eq.~(\ref{eq:EAAI_abbr}) can be replaced with
\begin{equation}
r_{ij}=\sum_{\forall(v^\alpha_a,v^\beta_b)\in \Phi} e^\alpha_{ia} \cdot h^\alpha_a \cdot e^\beta_{jb} + e^\alpha_{ia}\cdot h^\beta_b \cdot e^\beta_{jb}.
\label{eq:EAAI_withedge}
\end{equation}

If node $v^\alpha_a$ is a matched interlayer node in layer $\alpha$, a counterpart node must exist in layer $\beta$, and vice versa. Naturally, it is possible to make the PINPs uniform, as follows: $(v^\alpha_1$, $v^\beta_1),(v^\alpha_2$, $v^\beta_2),\cdots,(v^\alpha_n$, $v^\beta_n)$. Therefore, Eq.~(\ref{eq:EAAI_withedge}) can be rewritten as
\begin{equation}
r_{ij}=\sum_{a=1}^{n} h^\alpha_a \cdot e^\alpha_{ia} \cdot e^\beta_{aj} + e^\alpha_{ia} \cdot e^\beta_{aj} \cdot h^\beta_a.
\label{eq:EAAI_a}
\end{equation}

Using the vector form, Eq.~(\ref{eq:EAAI_a}) can be represented as
\begin{equation}
\begin{array}{l}

r_{ij}=[h^\alpha_1 \cdot e^\alpha_{i1} , h^\alpha_2 \cdot e^\alpha_{i2} , \cdots , h^\alpha_n \cdot e^\alpha_{in}] \cdot
\left[
 \begin{array}{c}
 e^\beta_{1j} \\
 e^\beta_{2j} \\
 \vdots\\
 e^\beta_{nj} \\
 \end{array}
\right]\\
+~[e^\alpha_{i1} , e^\alpha_{i2} , \cdots , e^\alpha_{in}] \cdot
\left[
 \begin{array}{c}
 e^\beta_{1j} \cdot h^\beta_1\\
 e^\beta_{2j} \cdot h^\beta_2\\
 \vdots\\
 e^\beta_{nj} \cdot h^\beta_n\\
 \end{array}
\right].
\end{array}
\label{eq:EAAI_vector}
\end{equation}

It is known that the Hadamard product of $A\equiv [a_{ij}]$ and $B\equiv [b_{ij}]$ with the same dimensions is the matrix $A\circ B=[a_{ij}b_{ij}]$. Therefore, $[h^\alpha_1 \cdot e^\alpha_{i1}, h^\alpha_2 \cdot e^\alpha_{i2}, \cdots, h^\alpha_n \cdot e^\alpha_{in}]$ can be represented as $[h^\alpha_1, h^\alpha_2, \cdots, h^\alpha_n] \circ [e^\alpha_{i1}, e^\alpha_{i2}, \cdots, e^\alpha_{in}]$.
Denoting $\bm{h}^\alpha_i=[h^\alpha_1, h^\alpha_2, \cdots, h^\alpha_n]^T$, $\bm{h}^\beta_j=[h^\beta_1, h^\beta_2, \cdots, h^\beta_n]^T$, $\bm{e}^\alpha_i=[e^\alpha_{i1}, e^\alpha_{i2}, \cdots, e^\alpha_{in}]^T$, $\bm{e}^\beta_j=[e^\beta_{1j}, e^\beta_{2j}, \cdots, e^\beta_{nj}]^T$, where $(\cdot)^T$ is the transposition of $(\cdot)$, Eq.~(\ref{eq:EAAI_vector}) can be rewritten as
\begin{equation}
r_{ij}=((\bm{h}^\alpha_i)^T \circ (\bm{e}^\alpha_i)^T ) \cdot \bm{e}^\beta_j + (\bm{e}^\alpha_i)^T \cdot (\bm{e}^\beta_j \circ \bm{h}^\beta_j).
\label{eq:EAAI_vector_sym}
\end{equation}

Using Eq.~(\ref{eq:EAAI_vector_sym}), the matching degree of UNP $(u^\alpha_i,u^\beta_j)$ is represented by the form of the vector operation. Then, we can express the matching degree of all UNPs in the matrix operation form. In a similar manner, we denote $\bm{H}^\alpha=[\bm{h}^\alpha_1,\quad \bm{h}^\alpha_2, \quad \cdots, \quad \bm{h}^\alpha_{n^\alpha-n}]$, $\bm{E}^\alpha=[\bm{e}^\alpha_1, \bm{e}^\alpha_2, \cdots, \bm{e}^\alpha_{n^\alpha-n}]$, $\bm{H}^\beta=[\bm{h}^\beta_1, \bm{h}^\beta_2, \cdots, \bm{h}^\beta_{n^\beta-n}]$, $\bm{E}^\beta=[\bm{e}^\beta_1, \bm{e}^\beta_2, \cdots, \bm{e}^\beta_{n^\beta-n}]$. The matching degree of all UNPs can be calculated as follows:
\begin{equation}
\bm{R}=([\bm{H}^\alpha]^T \circ [\bm{E}^\alpha]^T ) \cdot \bm{E}^\beta + [\bm{E}^\alpha]^T \cdot (\bm{E}^\beta \circ \bm{H}^\beta).
\label{eq:EAAI_all}
\end{equation}

It is worth noting that $\bm{H}^\alpha$ consists of $n^\alpha-n$ copies of $\bm{h}^\alpha_i$, because $\bm{h}^\alpha_1 = \bm{h}^\alpha_2 = \cdots = \bm{h}^\alpha_n=[h^\alpha_1, h^\alpha_2, \cdots, h^\alpha_n]^T$. Similarly, $\bm{H}^\beta$ consists of $n^\beta-n$ copies of $\bm{h}^\beta_j$. Moreover, $\bm{E}^\alpha$ and $\bm{E}^\beta$ are the submatrices of the adjacency matrix of $G^\alpha$ and $G^\beta$, respectively. Each row of the submatrix represents one of the matched interlayer nodes, while each column represents one of the unmatched nodes.

Based on the above statements, the four matrices $\bm{H}^\alpha$, $\bm{E}^\alpha$, $\bm{E}^\beta$, and $\bm{H}^\beta$ can be obtained. The matching degree of all unmatched nodes between the different layers of the multiplex network $\mathcal{M}$ can be calculated efficiently.
\subsection{Selecting matched interlayer node pairs}
After obtaining the values of all elements of the matching degree matrix $\bm{R}$, we need to determine which UNPs can be selected as MINPs. This means that we should formulate the objective function $J$. This function clarifies that a node pair can be selected as an MINP when its matching degree satisfies certain conditions. A larger value of the matching degree $r_{ij}$ indicates a higher probability that the node pair ($u^\alpha_i$, $u^\beta_j$) is the MINP. Therefore, we define the objective function as:
\begin{equation}
\begin{array}{c}
J(r_{ij})=\mathds{1}(r_{ij}\geq\delta \cdot \max(\bm{R})),\\
s.t. ~u^\alpha_i \notin \psi^\alpha, u^\beta_j \notin \psi^\beta\\
\end{array},
\label{eq:obj_func2}
\end{equation}
where (i) $\mathds{1}(\cdot)$ is an indicator function that takes 1 if the condition inside the parenthesis is true, and zero otherwise; (ii) $\max(\cdot)$ is the maximum function that takes the maximum value inside the parenthesis; (iii) $\psi^\alpha$ and $\psi^\beta$ are the sets of matched interlayer nodes in layers $\alpha$ and $\beta$, respectively; and (iv) $\delta$ is a control parameter that takes a value from 0 to 1. When $\delta=1$, only the UNPs with the maximum value of the matching degree can be selected as MINPs. When $\delta \in (0,1)$, the UNPs with a matching degree greater than $\delta \cdot \max(\bm{R})$ can be selected as MINPs. Moreover, when $\delta =0$, all of the UNPs can be selected as MINPs. It is obvious that a smaller value of $\delta$ means that additional UNPs are selected as MINPs, and hence, the accuracy is lower. In contrast, a larger value of $\delta$ indicates that less UNPs are selected as MINPs, and hence, the efficiency is lower. The constraints in Eq.~(\ref{eq:obj_func2}) are used to ensure that each unmatched node is matched only once.
\subsection{Achieving additional matched interlayer node pairs iteratively}
In the previous steps, the degree penalty principle and matrix operation are leveraged to calculate the matching degree matrix $\bm{R}$ for all unmatched nodes among the different layers in the multiplex network $\mathcal{M}$ and the node pairs with matching degrees greater than $\delta$ times $\max(\bm{R})$ are selected as MINPs. By performing these steps once, only one or several UNPs can be selected as MINPs. Therefore, we propose an iterative strategy, known as the IDP algorithm, to achieve additional MINPs.

Denoting $\Psi$ as the set of MINPs, we can add elements of $\Psi$ to $\Phi$. Moreover, we execute the steps introduced above again to identify more MINPs. This strategy can be executed iteratively until all of the unmatched nodes in one layer are matched or all of the matching degrees of the UNPs are equal to 0.

The details of our suggested IDP algorithm for interlayer link prediction are presented in Algorithm 1. In each iteration, the time complexity of the algorithm mainly depends on the calculation of the matching degree matrix $\bm{R}$ of Algorithm 1, which is matrix multiplication. Provided $n^\alpha-n=n^\beta-n=m$, the running time of the matrix multiplication is $nm^2/q$, where $q$ is the number of compute nodes~\cite{lee1997IO}. For the whole algorithm, provided that $z$ is the average number of selected matched interlayer nodes per iteration, the time complexity of IDP is $O(nm^3/zp)$. Moreover, an illustration of the procedure of the IDP algorithm is presented in Fig.~\ref{pic:IEAA_Procedure}.
\begin{table*}
\centering
\label{tab:algorithm}
\begin{tabular}{p{160mm}}
%\toprule
\hline
\textbf{Algorithm 1.} IDP\\
%\midrule
\hline
\textbf{Input:}Nodes and connections of $G^\alpha$ and $G^\beta$ in multiplex network $\mathcal{M}$, PINPs, parameter $\delta$.\\
\textbf{Output:}MINPs\\
\ 1: \textbf{function} IDP($G^\alpha$, $G^\beta$, PINPs, $\delta$)\\
\ 2:\quad $\bm{A}^\alpha \leftarrow$ adjacency matrix of $G^\alpha$, $\bm{A}^\beta \leftarrow$ adjacency matrix of $G^\beta$\\
\ 3:\quad $\bm{d}^\alpha \leftarrow$ degree of nodes in $G^\alpha$, $\bm{d}^\beta \leftarrow$ degree of nodes in $G^\beta$\\
\ 4:\quad $n^\alpha$=size($\bm{A}^\alpha$), $n^\beta$=size($\bm{A}^\beta$), $n$=size(PINPs)\\
\ 5:\quad  matchNode$^\alpha$=[], unmatchNode$^\alpha$=[], matchNode$^\beta$=[], unmatchNode$^\beta$=[], MINPs=[] \\
\ 6:\quad \textbf{while} $n<n^\alpha$ \textbf{and} $n<n^\beta$ \textbf{do}\\
\ 7:\quad \quad j=1,k=1\\
\ 8:\quad \quad \textbf{foreach} i \textbf{in} $n^\alpha$ \textbf{do}\\
\ 9:\quad \quad \quad \textbf{if} i \textbf{is in} PINPs \textbf{do}\\
10:\quad \quad \quad \quad matchNode$^\alpha$[j]=i, j++\\
11:\quad \quad \quad \textbf{else do}\\
12:\quad \quad \quad \quad unmatchNode$^\alpha$[k]=i, k++\\
13:\quad \quad j=1,k=1\\
14:\quad \quad \textbf{foreach} i \textbf{in} $n^\beta$ \textbf{do}\\
15:\quad \quad \quad \textbf{if} i \textbf{is in} PINPs \textbf{do}\\
16:\quad \quad \quad \quad matchNode$^\beta$[j]=i, j++\\
17:\quad \quad \quad \textbf{else do}\\
18:\quad \quad \quad \quad unmatchNode$^\beta$[k]=i, k++\\
19:\quad \quad $\bm{E}^\alpha$=submatrix($\bm{A}^\alpha$,matchNode$^\alpha$,unmatchNode$^\alpha$)\\
20:\quad \quad $\bm{E}^\beta$=submatrix($\bm{A}^\beta$,matchNode$^\beta$,unmatchNode$^\beta$)\\
21:\quad \quad $\bm{h}^\alpha_1$=1./log(submatrix($\bm{d}^\alpha$,matchNode$^\alpha$,1).+1)\\
22:\quad \quad $\bm{h}^\beta_1$=1./log(submatrix($\bm{d}^\beta$,matchNode$^\beta$,1).+1)\\
23:\quad \quad $\bm{H}^\alpha=[\bm{h}^\alpha_1,\bm{h}^\alpha_1,\cdots,\bm{h}^\alpha_1]_{n\times(n^\alpha-n)}$\\
24:\quad \quad $\bm{H}^\beta=[\bm{h}^\beta_1,\bm{h}^\beta_1,\cdots,\bm{h}^\beta_1]_{n\times(n^\beta-n)}$\\
25:\quad \quad $\bm{R}=((\bm{H}^\alpha)^T \circ (\bm{E}^\alpha)^T )\cdot \bm{E}^\beta + [\bm{E}^\alpha]^T \cdot (\bm{E}^\beta \circ \bm{H}^\beta)$\\
26:\quad \quad $\max=0$\\
27:\quad \quad \textbf{foreach} i \textbf{in} $n^\alpha-n$ \textbf{do}\\
28:\quad \quad \quad \textbf{foreach} j \textbf{in} $n^\beta-n$ \textbf{do}\\
29:\quad \quad \quad \quad \textbf{if} $\bm{R}$[i][j]$>\max$ \textbf{do}\\
30:\quad \quad \quad \quad \quad $\max=\bm{R}$[i][j]\\
31:\quad \quad \textbf{foreach} i \textbf{in} $n^\alpha-n$ \textbf{do}\\
32:\quad \quad \quad \textbf{foreach} j \textbf{in} $n^\beta-n$ \textbf{do}\\
33:\quad \quad \quad \quad \textbf{if} $\bm{R}$[i][j]$>=\delta \cdot \max$ \textbf{and} unmatchNode$^\alpha$[j] \textbf{not in} MINPs \textbf{do}\\
34:\quad \quad \quad \quad \quad MINPs $\leftarrow$ (unmatchNode$^\alpha$[i],unmatchNode$^\beta$[j])\\
35:\quad \quad \quad \quad \quad PINPs $\leftarrow$ (unmatchNode$^\alpha$[i],unmatchNode$^\beta$[j])\\
36:\quad \quad $n$=size(PINPs)\\
37:\quad \textbf{return} MINPs \\
38: \textbf{function} submatrix($\bm{A}$,rows,cols)\\
39:\quad  $\bm{Q}$ $\leftarrow$ Extracts the rows and columns of the matrix $\bm{A}$ according to the number in the arrays of rows and cols.\\
40:\quad  \textbf{return} $\bm{Q}$\\
%\bottomrule
\hline
\end{tabular}
\end{table*}
\section{Experiments}
In this section, we firstly describe the experimental settings. Thereafter, the evaluation metrics are introduced. Finally, we present the experimental results of the baseline and IDP algorithms on artificial scale-free networks and real-world networks.
\begin{table}[!h]
\caption{Time complexity of baselines and IDP}
\label{tab:timecomplexity}
\centering
\begin{tabular}{cc}
\hline
\textbf{Methods }& \textbf{Time Complexity}\\
\hline
INOE & $O(KD(|E^\alpha|+|E^\beta|))$\\
NS & $O(m(|E^\alpha|+|E^\beta|)d^{\alpha}d^{\beta})$\\
FRUI & $O(mnd^{\alpha}d^{\beta})$\\
IDP & $O(nm^3/zp)$\\
\hline
\end{tabular}
\end{table}
\subsection{Experimental settings}
We verify the effectiveness of our proposed IDP algorithm on both artificial scale-free and real-world networks. The steps for constructing the artificial scale-free multiplex networks are as follows:

(i) We create a Barab\'{a}si-Albett (BA)~\cite{Barabasi1999-BA} network, the degree distribution of which follows a power law distribution, according to the generation step proposed in Ref.~\cite{Barabasi1999-BA}. This is an original network for the following steps. (ii) Using $G^{org}(V^{org},E^{org})$ to represent the original BA network and $s$ to represent the percentage of remaining nodes, we construct two networks, $G^\alpha=G^{org}$ and $G^\beta=G^{org}$. (iii) For a node $v^\alpha_i$ in $G^\alpha$, we generate a random value $a_1$ with a uniform distribution in [0,1]. If $a_1\geq s$, the node $v^\alpha_i$ is discarded, as are all of the intralayer links connected with node $v^\alpha_i$. Otherwise, node $v^\alpha_i$ is preserved in $G^\alpha$. (iv) Similarly, for a node $v^\beta_i$ in $G^\beta$, we generate a random value $a_2$ with a uniform distribution in [0,1]. If $a_2\geq s$, the node $v^\beta_i$ is discarded, as are all of the intralayer links connected with node $v^\beta_i$. Otherwise, node $v^\beta_i$ is preserved in $G^\beta$. (v) We determine whether to add an interlayer link between $v^\alpha_i$ and $v^\beta_i$. If $a_1<s$ and $a_2<s$ simultaneously, the counterpart nodes of node $v^{org}_i$, namely $v^\alpha_i$ and $v^\beta_i$, remain in layers $\alpha$ and $\beta$. Therefore, we add an interlayer link between $v^\alpha_i$ and $v^\beta_i$. Otherwise, we do nothing. (vi) Steps (iii) to (v) are repeated until all of the nodes in $G^\alpha$ and $G^\beta$ are traversed.

The real-world datasets contain eight networks downloaded from the websites Stanford Large Network Dataset Collection~\cite{snapnets}, Link Prediction Group~\cite{linkpredictioncite}, and Network Analysis of Advogato~\cite{konect}. In a real scenario, OSN application users are often wary of their privacy. Therefore, it is difficult to obtain complete ground truth. To address this problem, we perform experiments on self-matching real-world networks\cite{KongChao2016}. The solution for obtaining the self-matching real-world multiplex network is the same as that for the BA multiplex network described above.

After constructing the multiplex networks, we use the FRUI~\cite{ZhouXiaoping2016}, NS~\cite{narayanan2009anonymizing}, and INOE~\cite{LiuLi2016} as the baselines for the experiments, where FRUI is the closest to the IDP algorithm for the interlayer link prediction as a state of the art. Each experiment is repeated 500 times. For the sake of reducing computational time in each experiment, the adjacency matrices constructed by the matched nodes and unmatched nodes of layer $\alpha$ and $\beta$ in the $t$th iteration will join to $\bm{E}^\alpha$ and $\bm{E}^\beta$, respectively. This could avoid reconstructing $\bm{E}^\alpha$ and $\bm{E}^\beta$ in $(t+1)$th iteration.

The time complexity of the baseline and IDP algorithms is presented in Table~\ref{tab:timecomplexity}. In particular, $K$ is the negative sampling number, $D$ is the representation dimension, $|E^\alpha|$ and $|E^\beta|$ are the number of intralayer links in layer $\alpha$ and $\beta$, respectively. $d^\alpha$ and $d^\beta$ are the maximal degrees of nodes in layer $\alpha$ and $\beta$, respectively.

\subsection{Evaluation metrics}
We employ the recall, precision, and F1 ~\cite{ShuKai2017} as the metrics for evaluating the performance of the FRUI and IDP algorithms, which are widely used in information retrieval, machine learning, and data mining, among others. The recall can be formulated as
\begin{equation}
Recall=\frac{TP}{TP+FN},
\end{equation}
where TP and FN indicate the number of true positives and false negatives, respectively~\cite{LiuJing2013-WSDM}. In this study, the recall reflects the ratio of INPs that the algorithm correctly predicts to the number of INPs that need to be predicted. The precision can be formulated as
\begin{equation}
Precision=\frac{TP}{TP+FP},
\end{equation}
where FP indicates the number of false positives~\cite{LiuJing2013-WSDM}. In this study, the precision reflects the ratio of INPs that the algorithm correctly predicts to all of the results predicted by the algorithm. $F1$ is the harmonic mean between the precision and recall~\cite{Hripcsak2005}. It can be formulated as
\begin{equation}
F1=\frac{2 \cdot Recall \cdot Precision}{Recall + Precision},
\end{equation}
where the range for $F1$ is [0,1]. For these three metrics, a higher value indicates superior performance of the algorithm performance. %It tells us how precise the algorithm is, as well as how robust it is.
\subsection{Results on artificial networks}
In this subsection, we outline the manner in which to determine an optimal value of $\delta$, and demonstrate the comparison results of the baseline and IDP algorithms on different average node degrees, node overlaps, and network sizes.
\subsubsection{Determining optimal $\delta$}
A possible solution for obtaining an optimum $\delta$ is to determine it experimentally. We set the default values of $\delta$ to increase from 0.1 to 1 by 0.1 and execute the IDP algorithm. In these experiments, the BA network parameter $m$, which represents the number of edges for attaching a new node to existing nodes, is equal to 10, the percentage of remaining nodes $s$ is equal to 0.5, the network sizes $N$ of the original BA networks are equal to 2,000,~4,000,~6,000,~8,000, and 10,000, respectively, and the ratios of PINPs to INPs $p$ are equal to $0.01,~0.02,~\cdots,~0.1$, respectively.

Table~\ref{tab:choose_delta} displays the average F1 rates of these experiments. It can be observed that the IDP algorithm exhibits the best performance at $\delta=0.5$ when $N$ is equal to 2,000. When $N$ is equal to $4,000,~6,000,~8,000, and~10,000$, the IDP algorithm exhibits superior performance at $\delta=0.5,~0.7,~0.7$, and $0.7$, respectively. Overall, the IDP algorithm performs the best when $\delta$ is equal to 0.7, at which the average F1 rate is equal to 0.3023. Therefore, we determine that the practical value of the parameter $\delta$ is equal to 0.7.
\begin{table*}%[htb]
\centering
\caption{Performance of IDP algorithm on different $\delta$}%Performance
\label{table}
\setlength{\tabcolsep}{3pt}
\begin{tabular}{cccccccccccc}
%\toprule
\hline
\multirow{2}*{\textbf{Metric}} & \multirow{2}*{$N$}
& \multicolumn{10}{c}{$\delta$} \\ \cline{3-12}
&& 0.1& 0.2& 0.3& 0.4& 0.5& 0.6& 0.7& 0.8& 0.9& 1.0\\
%\midrule
\hline
\multirow{6}*{$F_1$}
&2,000 &0.1029& 0.1113& 0.1355& 0.1435& \textbf{0.1445}& 0.1389& 0.1311& 0.1372& 0.1299& 0.1015\\
&4,000 &0.1786& 0.2043& 0.2303& 0.2462& 0.2548& \textbf{0.2574}& 0.2546& 0.2559& 0.2508& 0.2390\\
&6,000 &0.2199& 0.2515& 0.2881& 0.3073& 0.3176& 0.3287& 0.3331& 0.3294& \textbf{0.3344}& 0.3227\\
&8,000 &0.2445& 0.2868& 0.3207& 0.3398& 0.3505& 0.3665& 0.3711& 0.3674& \textbf{0.3727}& 0.3603\\
&10,000 &0.2786& 0.3207& 0.3554& 0.3771& 0.3998& 0.4133& 0.4217& 0.4180& \textbf{0.4221}& 0.4151\\
\cline{2-12}
&\textbf{Average} &0.2049& 0.2349& 0.2660& 0.2828& 0.2934& 0.3010& \textbf{0.3023}& 0.3016& 0.3020& 0.2877\\
%\bottomrule
\hline
\end{tabular}
\label{tab:choose_delta}
\end{table*}
\subsubsection{Effects of average degree}
\begin{figure*}%[t!]
\centering
 \includegraphics[width=1\textwidth]{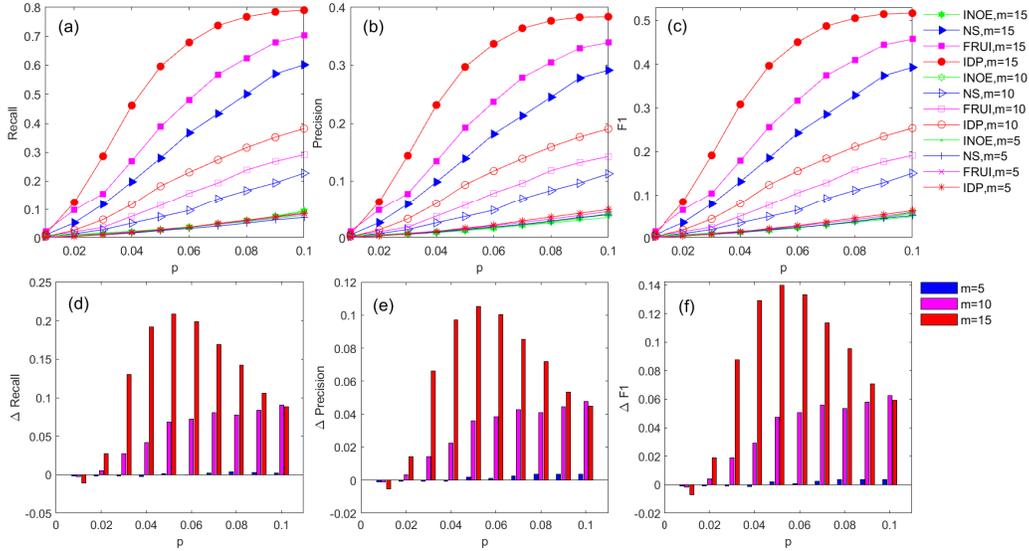}
 \caption{Comparison between baselines and IDP on different average degrees. (a) Recall rate, (b) precision rate, (c) F1 rate, (d) percentages of improvement between FRUI and IDP of recall rate, (e) percentages of improvement between FRUI and IDP of precision rate, and (f) percentages of improvement between FRUI and IDP of F1 rate versus $p$. In these experiments, we set $N=2,000$, $s=0.5$, and $m=5,~10,~15$. The horizontal ordinates denote the ratios of PINPs to INPs.}
 \label{fig:BA-mChg}
\end{figure*}

We evaluate the performance of the baseline and IDP algorithms at different average degrees. We set $N=2,000$, $s=0.5$, $p$ increasing from 0.01 to 0.1 by 0.01, and $m=5,~10,~15$. Figures~\ref{fig:BA-mChg}(a) to (c) display the recall, precision, and F1 rates of the baseline and IDP algorithms under the above experimental settings. The following observations can be made. The IDP algorithm outperforms the baseline algorithms. This is because the IDP algorithm uses the degree penalty principle to calculate the matching degree of two unmatched nodes across different layers. This principle can reflect the contribution of the number of CMNs and influence of the degree of CMNs on the matching degree, while the FRUI algorithm only reflects the contribution of the number of CMNs. The NS algorithm depends heavily on the degree of the nodes in layer $\beta$. It limits the contribution of CMNs; hence, the performance is not as good as FRUI. The INOE algorithm uses network embedding to predict the interlayer links, which needs more PINPs to train the model to show its advantage. For a given $m$, the recall, precision, and F1 rates of the baseline and IDP algorithms increase with $p$. This is because a larger $p$ means that additional CMNs can be used to calculate the matching degree of UNPs. For a given $p$, the recall, precision, and F1 rates of the baseline and IDP algorithms increase with $m$, as the nodes with a low degree are difficult to match. A smaller $m$ indicates more low-degree nodes. It is noteworthy that some lines in Figs.~\ref{fig:BA-mChg}(a) to (c) are overlapped. We have listed their recall rate in Table~\ref{tab:recall_overlap}.
\begin{table*}%[htb]
\centering
\caption{Recall rate of overlapped lines in Figs.~\ref{fig:BA-mChg}(a) to (c)}%Performance
\label{table}
\setlength{\tabcolsep}{3pt}
\begin{tabular}{cccccccccccc}
%\toprule
\hline
\multirow{2}*{\textbf{Method}} & \multirow{2}*{$m$}
& \multicolumn{10}{c}{$p$} \\ \cline{3-12}
&& 0.01& 0.02& 0.03& 0.04& 0.05& 0.06& 0.07& 0.08& 0.09& 0.1\\
%\midrule
\hline
INOE&15 &0.0079& 0.0108& 0.0147& 0.0200& 0.0273& 0.0365& 0.0481& 0.0610& 0.0764& 0.0950\\
INOE&10 &0.0074& 0.0106& 0.0153& 0.0215& 0.0289& 0.0383& 0.0489& 0.0608& 0.0752& 0.0912\\
INOE&5 &0.0084& 0.0123& 0.0176& 0.0240& 0.0307& 0.0386& 0.0475& 0.0574& 0.0680& 0.0795\\
NS&5 &0.0037& 0.0071& 0.0113& 0.0174& 0.0260& 0.0338& 0.0414& 0.0521& 0.0633& 0.0718\\
FRUI&5 &0.0038& 0.0077& 0.0134& 0.0213& 0.0277& 0.0385& 0.0489& 0.0594& 0.0713& 0.0843\\
IDP&5 &0.0022& 0.0060& 0.0117& 0.0192& 0.0293& 0.0382& 0.0509& 0.0627& 0.0743& 0.0866\\
%\bottomrule
\hline
\end{tabular}
\label{tab:recall_overlap}
\end{table*}

Figures~\ref{fig:BA-mChg}(d) to (f) illustrate the percentages of improvement in the recall, precision, and F1 rates for the IDP algorithm compared to the best baseline, FRUI, algorithm under the same experimental settings as those in Figs.~\ref{fig:BA-mChg}(a) to (c). The recall rate increases by a maximum of 20.8\% and an average of 6.0\%. The precision rate increases by a maximum of 10.5\% and an average of 3.1\%. The F1 rate increases by a maximum of 14\% and an average of 4.1\%. A larger $m$ results in greater overall improvement percentages. This is because, with a larger $m$, each unmatched node or matched interlayer node has more interlayer links. For a UMP, more CMNs may be involved in the calculation of its matching degree. Thus, the advantage of the degree penalty principle becomes more obvious. When $m=15$, the percentages of improvement exhibit a trend of first increasing and then decreasing with an increase in $p$. This trend is caused by the following factors. (i) When $p$ is very small, the number of PINPs used to calculate the matching degree is small. The ability of both the FRUI and IDP algorithms is poor. Thus, the percentages of improvement are not obvious. (ii) With an increase in $p$, the number of PINPs increases. Owing to the use of the degree penalty principle to calculate the matching degree, the IDP algorithm can use more useful information of the CMNs. A greater number of PINPs results in higher percentages of improvement. (iii) When the percentages of improvement reach a maximum value, the matching advantage offered by increasing the PINPs in the IDP algorithm decreases. Thus, the percentages of improvement gradually decrease until they disappear. When $m=5$ or $m=10$, the percentages of improvement only exhibit an increasing trend. This trend is consistent with the first half when $m=15$. When $p$ is greater than 0.1 and the percentages of improvement reach the maximum value, they will decrease as $p$ increases.
\subsubsection{Effects of node overlaps}
We use the Jaccard coefficient to measure the node overlaps, as follows:
\begin{equation}
O(G^\alpha,G^\beta)=\frac{|V^\alpha \bigcap V^\beta|}{|V^\alpha \bigcup V^\beta|}.
\label{eq:overlap}
\end{equation}
When $s=0.4,~0.5,~0.6$, the node overlapping rate is approximately equal to 0.25, 0.33, and 0.43, respectively, according to Eq.~(\ref{eq:overlap}).
\begin{figure*}[t!]
\centering
 \includegraphics[width=1\textwidth]{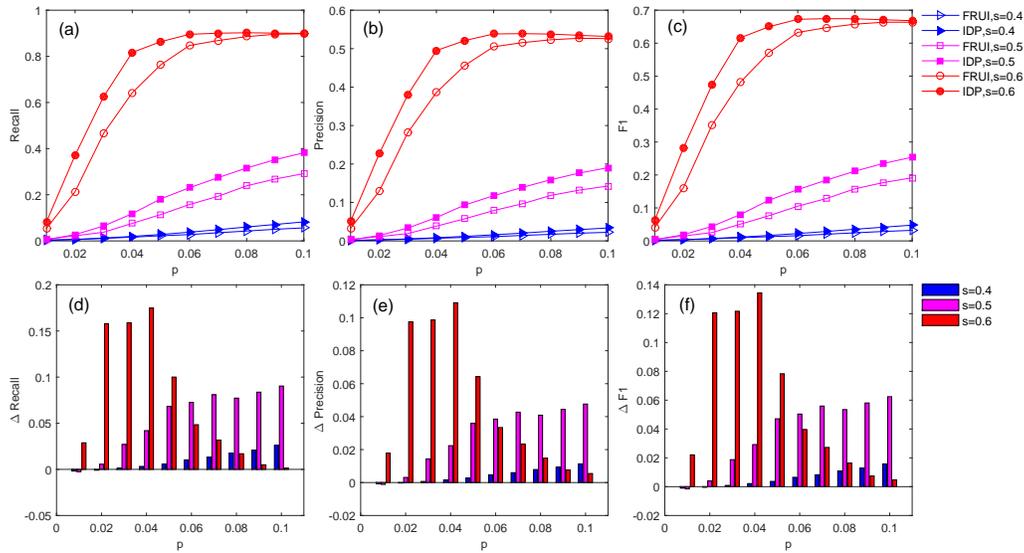}
 \caption{Comparison between FRUI and IDP algorithms on different node overlaps. (a) Recall rate, (b) precision rate, (c) F1 rate, (d) percentages of improvement of recall rate, (e) percentages of improvement of precision rate, and (f) percentages of improvement of F1 rate versus $p$. In these experiments, we set $N=2,000$, $m=10$, and $s=0.4,~0.5,~0.6$. The horizontal ordinates denote the ratios of PINPs to INPs.}
 \label{fig:BA-sChg}
\end{figure*}

We set $N=2000$, $m=10$, $p$ increasing from 0.01 to 0.1 by 0.01, and $s=0.4,~0.5,~0.6$ to execute the experiments for the evaluation of the performance of the FRUI and IDP algorithms with different node overlaps. Figures~\ref{fig:BA-sChg}(a) to (c) illustrate the recall, precision, and F1 rates of the FRUI and IDP algorithms under the above experimental settings. The following observations can be made.

The IDP algorithm outperforms the FRUI algorithm. For a given $s$, the recall, precision, and F1 rates of the FRUI and IDP algorithms increase with $p$. The reasons are the same as those in Figs.~\ref{fig:BA-mChg}(a) to (c). For a given $p$, the recall, precision, and F1 rates of the FRUI and IDP algorithms increase with $s$. This is because a higher overlapping rate results in a higher proportion of interlayer nodes in all nodes, and a lower probability of incorrect matching; hence, superior performance of the algorithms. For a UMP, additional interlayer links mean that more CMNs may be involved in the calculation of its matching degree. Thus, the advantage of the degree penalty principle becomes more obvious.

Figures~\ref{fig:BA-sChg}(d) to (f) display the percentages of improvement under the same experimental settings as those in Figs.~\ref{fig:BA-sChg}(a) to (c). The recall rate increases by a maximum of 17.5\% and an average of 4.6\%. The precision rate increases by a maximum of 10.9\% and an average of 2.7\%. The F1 rate increases by a maximum of 13.4\% and an average of 3.4\%. A larger $s$ results in greater overall percentages of improvement. This is because a larger $s$ indicates a greater number of INPs and PINPs. For a UMP, additional CMNs may be involved in the calculation of its matching degree; thus, the advantage of the degree penalty principle is more obvious. When $s=0.6$, the percentages of improvement exhibit a trend of first increasing and then decreasing with an increase in $p$. Moreover, when $s=0.4$ or $s=0.5$, the percentages of improvement exhibit a trend of increasing with an increase in $p$. The reasons are the same as those in Figs.~\ref{fig:BA-mChg}(d) to (f).
\subsubsection{Effects of network size}
\begin{figure*}[t!]
\centering
 \includegraphics[width=1\textwidth]{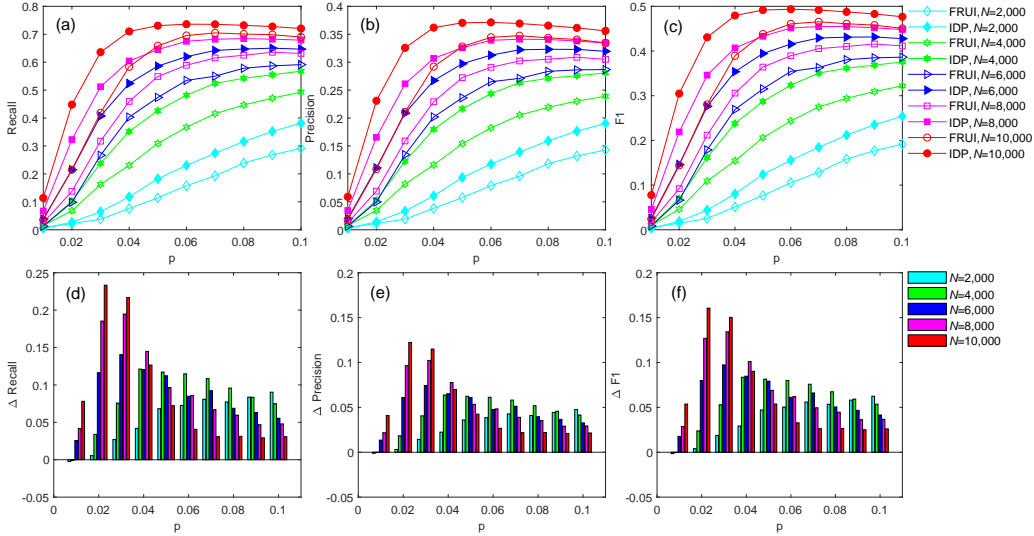}
 \caption{Comparison between FRUI and IDP algorithms on different number of nodes. (a) Recall rate, (b) precision rate, (c) F1 rate, (d) percentages of improvement of recall rate, (e) percentages of improvement of precision rate, and (f) percentages of improvement of F1 rate versus $p$. In these experiments, we set $m=10$, $s=0.5$, and $N=2,000,~4,000,~6,000,~8,000$, and 10,000. The horizontal ordinates denote the ratios of PINPs to INPs.}
 \label{fig:BA-nChg}
\end{figure*}

We set $s=0.5$, $m=10$, $p$ increasing from 0.01 to 0.1 by 0.01, and  $N=$2,000, 4,000, 6,000, 8,000, and 10,000 to execute the experiments for the evaluation of the performance of the FRUI and IDP algorithms with different network sizes. Figures~\ref{fig:BA-nChg}(a) to (c) illustrate the recall, precision, and F1 rates of the FRUI and IDP algorithms under the above experimental settings. The following observations can be made.

The IDP algorithm outperforms the FRUI algorithm. For a given $N$, the recall, precision, and F1 rates of the FRUI and IDP algorithms increase with $p$. The reasons are the same as those in Figs.~\ref{fig:BA-mChg}(a) to (c). For a given $p$, the recall, precision, and F1 rates of the FRUI and IDP algorithms increase with $N$. This is because, when $N$ is larger, more PINPs are used to calculate the matching degree of all UNPs; hence, the performance of the algorithms is improved. When $N=10,000$ and $p\geq0.07$, the performance of both the FRUI and IDP algorithms decreases slightly with the increase in $p$, because the number of interlayer links that can be predicted correctly has a maximum. When the maximum number is reached, the increase in $p$ will lead to a slight decrease in the number of interlayer links to be predicted and that are correctly predicted.

Figures~\ref{fig:BA-nChg}(d) to (f) display the percentages of improvement under the same experimental settings as those in Figs.~\ref{fig:BA-nChg}(a) to (c). The recall rate increases by a maximum of 23.3\% and an average of 8.2\%. The precision rate increases by a maximum of 12.2\% and an average of 4.5\%. The F1 rate increases by a maximum of 16.1\% and by an average of 5.8\%. A larger $N$ results in greater average percentages of improvement for the IDP algorithm compared to the FRUI algorithm. Because the BA networks exhibit the characteristics of preferential attachment, new vertices attach preferentially to vertices that are already well connected~\cite{Barabasi1999-BA}. A larger $N$ means that the influence of the degree of CMNs on the matching degree is more sensitive; hence, the percentages of improvement are increased. When $N=4,000,~6,000,~8,000$ or 10,000, the percentages of improvement exhibit a trend of first increasing and then decreasing with an increase in $p$. Moreover, when $N=2000$, the percentages of improvement exhibit a trend of increasing with an increase in $p$. The reasons are the same as those in Figs.~\ref{fig:BA-mChg}(d) to (f).

\subsection{Results on real-world networks}
\begin{table*}%[ht]
\centering
\caption{Statistical characteristics of eight real-world networks, including network size ($N$), number of edges ($E$), maximum degree ($k_{max}$), first ($\langle{k}\rangle$) and second moments ($\langle{k^2}\rangle$) of degree distribution, degree-degree correlations ($r$), and clustering ($c$).}
\setlength{\tabcolsep}{3pt}
\begin{tabular}{ccccccccc}
%\toprule
\hline
\multirow{2}*{\textbf{No.}}&\multirow{2}*{\textbf{Networks}}
&\multicolumn{7}{c}{\textbf{Statistical characteristics of networks}}\\ \cline{3-9}
&& $N$& $E$& $k_{max}$& $\langle{k}\rangle$ & $\langle{k^2}\rangle$& $r$& $c$\\
%\midrule
\hline
Real-1 & Email-Eu-core\cite{YinHao2017}& 1005& 16064& 345& 32.58& 2386.6 & $-0.026$& 0.45 \\
Real-2& UC Irvine messages\cite{opsahl2009clustering}& 1899& 13838& 255& 14.57& 810.7 & $-0.188$& 0.14 \\
Real-3&Wikipedia vote\cite{Leskovec2010}& 7115& 100762&1065 & 28.32& 3530.5 & $-0.083$& 0.21\\
Real-4&Twitter\cite{ZhangJiawei2015-IJCAI}& 5120& 130576& 1725& 51.01& 11679.7 & $-0.214$& 0.30 \\
Real-5&Political blogs\cite{AdamicLada2005}& 1222& 16714& 351& 27.36& 2223& $-0.221$& 0.36 \\
Real-6&Hamsterster friendships\cite{Kunegis2013}& 1788& 12476 & 272 &13.96 &635.6 &$-0.089$ &0.17 \\
Real-7&Hamsterster full\cite{Kunegis2013}& 2000& 16098& 273& 16.1 &704.7 &0.023 & 0.57\\
Real-8&Foursquare\cite{ZhangJiawei2015-IJCAI}& 5313& 54233& 552& 20.42& 1436.1 & $-0.193$& 0.23\\
%\bottomrule
\hline
\end{tabular}
\label{tab:real_statical}
\end{table*}
We use eight real-world networks to construct self-matching real-world multiplex networks, and evaluate the FRUI and IDP algorithms on these multiplex networks. The statistical characteristics of the eight real-world networks are presented in Table~\ref{tab:real_statical}. Formally, we represent each of the eight real-world networks as an undirected graph. In these experiments, $s=0.5$ and $p$ increases from 0.01 to 0.1 by 0.01.

\begin{figure*}%[t!]
\centering
 \includegraphics[width=1\textwidth]{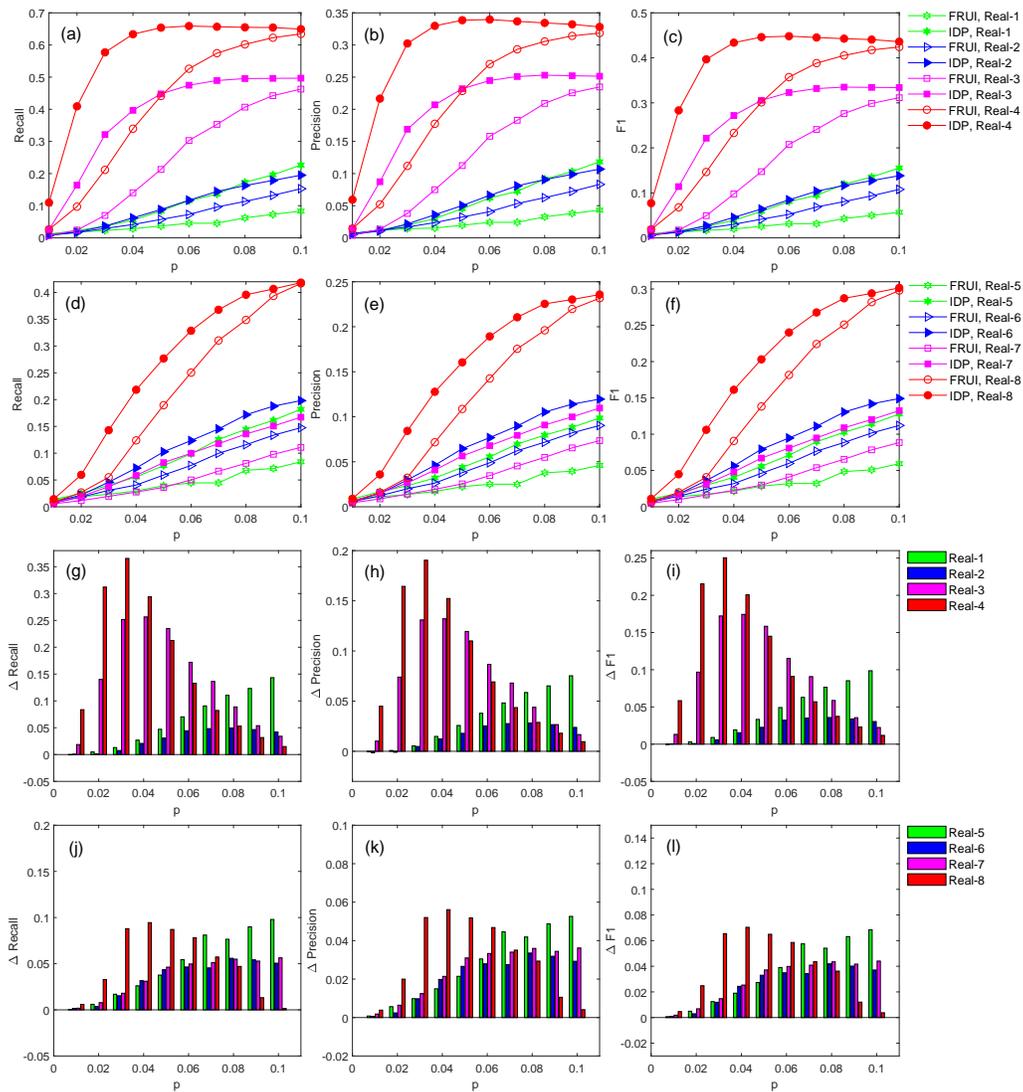}
 \caption{Comparison between FRUI and IDP algorithms on eight real-world networks. (a) and (d) Recall rate, (b) and (e) precision rate, (c) and (f) F1 rate, (g) and (j) percentages of improvement of recall rate, (h) and (k) percentages of improvement of precision rate, (i) and (l) percentages of improvement of F1 rate versus $p$. In these experiments, $s=0.5$. The horizontal ordinates denote the ratios of PINPs to INPs.}
 \label{fig:BA-real}
\end{figure*}

Figures ~\ref{fig:BA-real}(a) to (f) display the recall, precision, and F1 rates of the FRUI and IDP algorithms of the eight real-world networks. The following observations can be made. The IDP algorithm outperforms the FRUI algorithm. For a given real-world network, the recall, precision, and F1 rates of the FRUI and IDP algorithms increase with $p$. The reasons are the same as those in Figs.~\ref{fig:BA-mChg}(a) to (c). For a given $p$, different real-world networks exhibit varying performances, because the network sizes, average degrees, and clustering coefficients of the real-world networks differ.

Figures ~\ref{fig:BA-real}(g) to (l) illustrate the percentages of improvement under the same experimental settings as those in Figs.~\ref{fig:BA-real}(a) to (f). The following observations can be made. The recall rate increases by a maximum of 36.6\% and an average of 7.0\%. The precision rate increases by a maximum of 19.0\% and an average of 3.8\%. The F1 rate increases by a maximum of 25.0\% and an average of 5.0\%. On real-world network 2, 3, 4, and 8, the percentages of improvement exhibit a trend of first increasing and then decreasing with an increase in $p$. On real-world networks 1, 5, 6, and 7, the percentages of improvement exhibit a trend of first increasing and then decreasing with an increase in $p$. The reasons are the same as those in Figs.~\ref{fig:BA-mChg}(d) to (f).
\section{Conclusion}
In this study, we have investigated the problem of interlayer link prediction in the multiplex network. We solved this problem by leveraging network structure attributes. We used a degree penalty principle to calculate the matching degree of two unmatched nodes across different layers, which could reflect the influence of the number of CMNs and their degree attributes. For the sake of efficiency, we adopted node adjacency matrix multiplication to obtain the matching degrees of all UNPs. Moreover, we developed an iterative algorithm to determine additional hidden interlayer links. Experiments on both artificial and real-world networks demonstrated that our advanced IDP algorithm outperformed the FRUI algorithm.

In summary, the IDP algorithm offers comparative advantages in multiplex networks with a degree distribution that follows a power law distribution. When users in multiple OSN applications input different usernames and other attribute information owing to privacy or anonymity, our proposed algorithm can effectively associate the accounts belonging to the same person across different OSNs. Such an association can aid in achieving the information fusion of multiple OSN platforms, understanding the online behavior of network users, constructing integral user profiles, and providing services for e-commerce, cyber security, and recommendation systems, among others. As future work, we will investigate the manner in which to use higher-order structures, such as communicating within a group or participating in the same online activity, to improve the performance of the interlayer link prediction.

\section{Acknowledgments}
This work was supported by the National Science and Technology Support Program of China (Grant No. 2012BAH18B05) and the National Natural Science Foundation of China (Grant Nos. 81602935, 81773548, 61802270, 61802271).

%%% Loading bibliography style file
%%\bibliographystyle{cas-model2-names}
%
%% Loading bibliography database
%\bibliography{References}

\end{document}